\documentclass{emulateapj}

\usepackage{amsmath}
\usepackage{amssymb}
\usepackage{amsfonts}
\usepackage{amstext}
\usepackage{graphicx}
\usepackage{fancybox}
\usepackage{enumitem}
\usepackage[usenames,dvipsnames]{color}
\usepackage{pstricks}
\usepackage{bm}
\usepackage{soul, color}
\usepackage{latexsym}
\usepackage{pifont}
\usepackage{shorttoc}
\usepackage{subfigure}
\usepackage{textcomp} 
\usepackage{hyperref}
\usepackage{bm}
\usepackage{lipsum}
\usepackage{threeparttable}
\usepackage{epsfig}

\def\aap{A\&A}
\def\mnras{MNRAS}
\def\apj{ApJ}
\def\apjs{ApJS}
\def\apjl{ApJL}
\def\nat{Nature}

\def\ssr{SSR}

\newcommand{\cm}{\, {\rm cm}}

\newcommand{\no}[1]{}

\newcommand{\hs}{\hspace{1mm}}

\newcommand{\fescion}{$f^{\rm ion}_{\rm esc}$}
\newcommand{\fescionmx}{$f^{\rm ion}_{\rm esc,max}$}
\newcommand{\fesclya}{$f^{{\rm Ly}\alpha}_{\rm esc}$}
\newcommand{\HI}{{\text{H\MakeUppercase{\romannumeral 1}}} }
\newcommand{\Lya}{\ifmmode{{\rm Ly}\alpha}\else Ly$\alpha$\ \fi}

\def\lsim{~\rlap{$<$}{\lower 1.0ex\hbox{$\sim$}}}

\def\gsim{~\rlap{$>$}{\lower 1.0ex\hbox{$\sim$}}}

\shorttitle{The Ly$\alpha$-LyC Connection}
\shortauthors{Dijkstra et al.}

\begin{document}

\title{The Ly$\alpha$-LyC Connection: Evidence for an Enhanced Contribution of UV-faint Galaxies to Cosmic Reionization}

\author{Mark Dijkstra$^{1}$, Max Gronke$^{1}$ \& Aparna Venkatesan$^{2}$}
\affil{$^1$Institute of Theoretical Astrophysics, University of Oslo,
P.O. Box 1029 Blindern, N-0315 Oslo, Norway}
\affil{$^2$Department of Physics and Astronomy, University of San Francisco, 2130 Fulton Street, San Francisco, CA
94117I}
\altaffiltext{1}{mark.dijkstra@astro.uio.no}

\begin{abstract}
The escape of ionizing Lyman Continuum (LyC) photons requires the existence of low-$N_{\rm HI}$ sightlines, which also promote escape of Lyman-Alpha (Ly$\alpha$). We use a suite of $2500$ Ly$\alpha$ Monte-Carlo radiative transfer simulations through models of dusty, clumpy interstellar (`multiphase') media from Gronke \& Dijkstra (2016), and compare the escape fractions of Ly$\alpha$ (\fesclya) and LyC radiation (\fescion). We find that \fescion \hs and \fesclya \hs are correlated: galaxies with a low \fesclya \hs consistently have a low \fescion, while galaxies with a high \fesclya \hs exhibit a large dispersion in \fescion.  We argue that there is increasing observational evidence that Ly$\alpha$ escapes more easily from UV-faint galaxies. The correlation between \fescion \hs and \fesclya \hs then implies that UV-faint galaxies contribute more to the ionizing background than implied by the faint-end slope of the UV-luminosity function. In multiphase gases, the ionizing escape fraction is most strongly affected by the cloud covering factor, $f_{\rm cl}$, which implies that \fescion\hs is closely connected to the observed Ly$\alpha$ spectral line shape. Specifically, LyC emitting galaxies typically having narrower, more symmetric line profiles. This prediction is qualitatively similar to that for `shell models'.
\end{abstract}


 
\section{Introduction}
\label{sec:intro}

The escape fraction of ionizing photons, \fescion, represents one of the key parameters describing cosmic reionization \citep[e.g.][]{HH03,Cen03,WL03,Mitra}. Observational constraints on \fescion\hs are still weak \citep[see Fig~13 of][]{Smith16}. Ionizing photons, also known as Lyman Continuum (LyC) photons, have only been directly observed to escape for a handful of galaxies \citep[e.g.][also see Benson et al. 2013, Smith et al. 2016 and references therein]{B14,Izotov16,V16a}. Observations of the Ly$\alpha$ forest constrain the LyC volume emissivity (the rate at which LyC photons are {\it released} into the IGM per unit volume), while observations of the UV-luminosity function of star forming galaxies provide direct constraints on the {\it production} rate of LyC photons. These two constraints combined constrain the volume-averaged escape fraction of ionizing photons, denoted with $\langle$\fescion$\rangle$, and show that $\langle$\fescion$\rangle$ increases with redshift \citep[][]{Inoue06,Kuhlen,BB13}. 

The LyC escape fraction depends on more than just redshift. Various models and simulations predict that \fescion\hs decreases with dark matter halo mass \citep[e.g.][but also see Gnedin et al. 2008, Ma et al. 2015, Sharma et al. 2016]{Yajima11,FL13,P13,Wise14}, which in turn correlates with observables such as the non-ionizing UV-continuum luminosity of galaxies. The reason that not all simulations agree on this mass-dependence is partly because different studies focus on galaxies with very different masses, at very different redshifts, and different implementations for sub-grid physics associated with feedback, which can strongly affect the properties of the simulated interstellar medium. Ab initio modeling of \fescion\hs still represents a major theoretical challenge (see e.g. Fernandez \& Shull 2011 for a discussion), and models may have to include additional physical processes  such as X-ray heating/ionization (Benson et al. 2013), runaway stars \citep{CK12} and binary evolution (Ma et al. 2016), all of which can facilitate the escape of ionizing photons.

Irrespective of theoretical and observational uncertainties, the escape of ionizing photons requires that paths exist which contain low column densities of atomic hydrogen, i.e. $N_{\rm HI} \lsim 1/\sigma_{\rm ion} \approx 10^{17}$ cm$^{-2}$, where $\sigma_{\rm ion}=6\times 10^{-18}$ cm$^{2}$ denotes the photoionization cross-section evaluated at the Lyman limit \citep[e.g.][]{Verner96}. These same low column density paths provide escape routes for Ly$\alpha$ photons (Behrens et al. 2014, Verhamme et al. 2015). LyC and Ly$\alpha$ escape are therefore expected to be correlated, at least at some level (e.g. Rauch et al. 2011, Erb et al. 2014, Micheva et al. 2016). If the escape of LyC photons is facilitated by (supernova-driven) winds that blew low-column density holes (see e.g. Dove et al. 2000, Sharma et al. 2016), then this provides a physical mechanism connecting \fescion \hs and \fesclya, as observations of Ly$\alpha$ emitting galaxies indicate that galactic outflows promote the escape of Ly$\alpha$ photons (Kunth et al. 1998, Atek et al. 2008, Wofford et al. 2013, Rivera-Thorsen et a. 2015, see Hayes 2015 for a review).

The goal of this paper is to more quantitatively explore the correlation between Ly$\alpha$ and LyC photons, for which we use a large suite of simplified models of the multi-phase ISM that span the wide range of physical conditions encountered in observed galaxies (first presented in Gronke \& Dijkstra 2016). Yajima et al. (2014) previously found a clear correlation between \fesclya\hs and \fescion\hs in their cosmological hydrodynamical simulations of a Milky Way-like galaxy. Their calculations should be viewed as a `bottom-up'  (or ab-initio) approach to quantifying this correlation, while our work should be viewed as a `top-down' (or empirical) approach. As neither approach has converged yet (see \S~\ref{sec:model}), our work should be viewed as complementary to that of Yajima et al. (2014). Addressing the correlation between \fesclya\hs and \fescion\hs has become (even) more relevant for cosmic reionization as, we will argue in \S~\ref{sec:muvfesc},  there is increasing evidence that \fesclya\hs increases towards lower UV-luminosities.

The outline of this paper is as follows: In \S~\ref{sec:model} we present our models and show the predicted correlation between \fesclya \hs and \fescion\hs in \S~\ref{sec:results}. We discuss implications of our results in \S~\ref{sec:discuss}, before presenting our conclusions in \S~\ref{sec:conc}.

\section{The Model}
\label{sec:model}

\begin{figure*}
\centering
\includegraphics[width=8.5cm,angle=0]{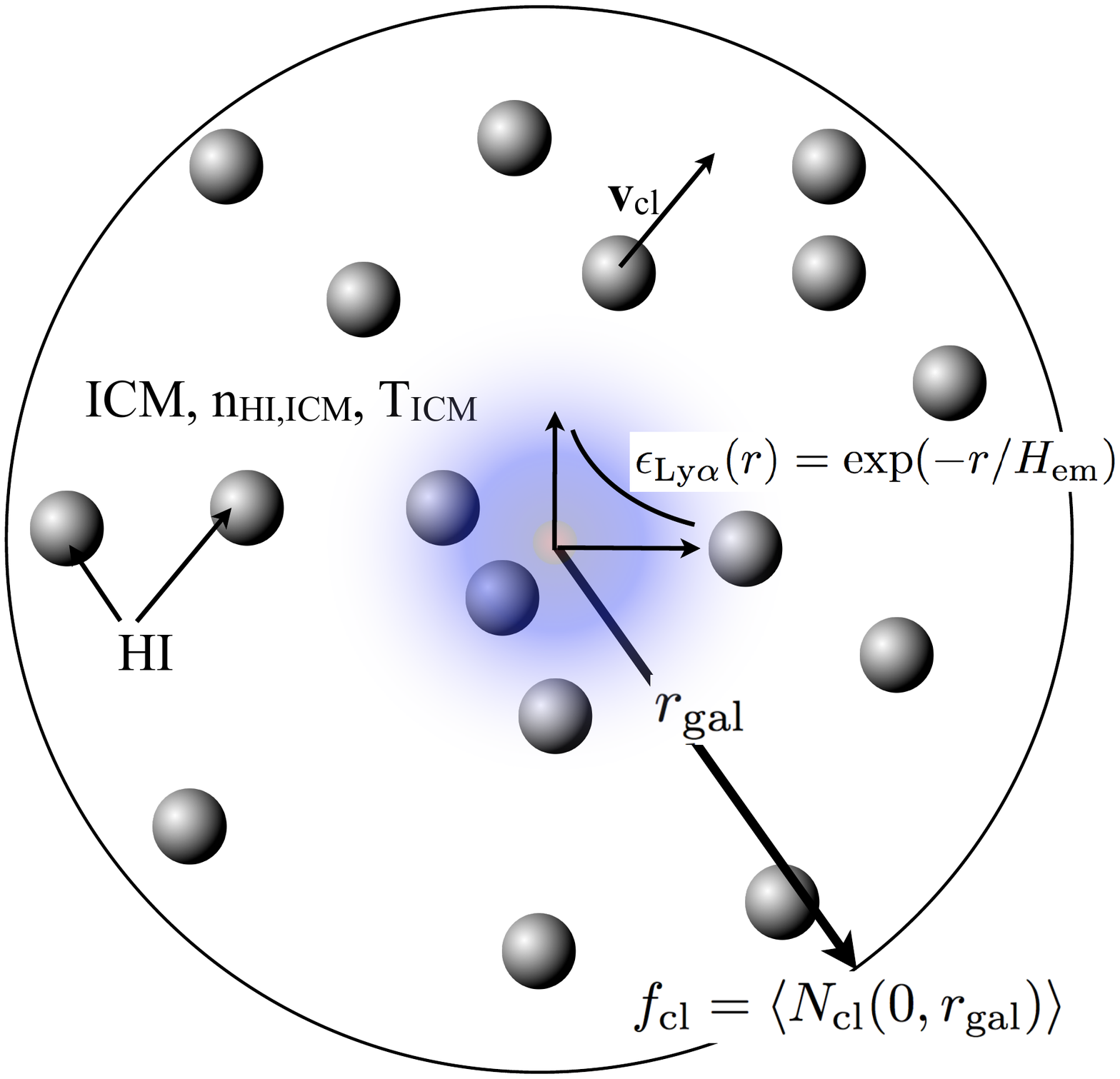}
\vspace{0mm}
\caption[]{Schematic representation of the adopted geometry in our `clumpy ISM' models, which represent simplified versions of multiphase interstellar media. A sphere or radius $r_{\rm cl}=5$ kpc is filled with outflowing, neutral, dusty clumps of gas embedded within a hot interclump medium. The covering factor $f_{\rm cl}$ denotes the average number of clumps along sightlines from the center to the edge of the cloud. The clumps surround a spatially extended Ly$\alpha$ (and LyC) source, both of which are characterized by an exponential volume emissivity profile with scale length $H_{\rm em}$. A fraction $P_{\rm cl}$ of all Ly$\alpha$ and LyC photons is emitted inside cold clumps.}
\label{fig:scheme}
\end{figure*} 

The escape of both ionizing and Ly$\alpha$ photons depends sensitively on the distribution of neutral gas throughout the interstellar medium. For Ly$\alpha$ photons, the kinematics of this neutral gas is possibly even more important (Kunth et al. 1998, Atek et al. 2008, Steidel et al. 2010, Wofford et al. 2013, Rivera-Thorsen et al. 2015). Modeling Ly$\alpha$ transfer on interstellar scales therefore requires a proper model for both the distribution and kinematics of the neutral gas in the ISM, which likely requires magneto-hydrodynamical simulations with sub-pc resolution (e.g. Fujita et al. 2009, Dijkstra \& Kramer 2012). This requirement underlines why it is important to have a complementary top-down approach to the bottom-up analysis by Yajima et al. (2014), whose simulations had a spatial resolution of $250h^{-1}$ comoving pc and gas mass resolution of $M=3\times 10^5h^{-1}M_{\odot}$. 

To circumvent the demanding requirements to properly model interstellar Ly$\alpha$ transfer from first principles, this process has been represented by highly simplified models, which include ({\it i}) the `shell' model, which consists of a Ly$\alpha$ source surrounded by a geometrically thin shell of neutral, dusty hydrogen, which is (typically) outflowing (see e.g. Ahn et al. 2003, Verhamme et al. 2006, Gronke et al. 2015a). The shell model -which contains seven free parameters - has been remarkably successful at reproducing observed Ly$\alpha$ spectra line profiles (e.g. Verhamme et al. 2008, Hashimoto et al. 2015, Yang et al. 2016, though some issues have been pointed out by Barnes \& Haehnelt 2010, Kulas et al. 2012, Chonis et al. 2013); and ({\it ii}) the `clumpy ISM' model, which consists of a (large) collection of spherical clumps that contain dusty, neutral hydrogen gas, embedded within a hot inter-clump medium, and which represent simplified versions of multiphase interstellar media (e.g. Neufeld 1991, Hansen \& Oh 2006, Laursen et al. 2013, Gronke \& Dijkstra 2014). Clumpy models naturally give rise to a non-zero porosity of the neutral gas, and a `continuum covering factor' of neutral gas that is less than $100\&$, both of which facilitates Ly$\alpha$ escape (e.g. Shibuya et al. 2014, Trainor et al. 2015, Rivera-Thorsen et al. 2015). Both sets of simplified models can be interpreted as `sub-grid' models that describe the Ly$\alpha$ transfer on scales that have not been modelled yet from first principles.

In shell models, the shell completely surrounds the Ly$\alpha$ source. The escape fraction \fescion\hs is determined by its HI column ($N_{\rm HI}$) as \fescion$(\nu)=\exp[-\sigma_{\rm ion}(\nu)N_{\rm HI}]$, and \fescion\hs is practically binary (\fescion$\approx0$ for $N_{\rm HI} > 10^{17}$ cm$^{-2}$ or \fescion$\approx1.0$ for $N_{\rm HI} \lsim 10^{17}$ cm$^{-2}$). However the production rate of Ly$\alpha$ is zero for \fescion$\approx 1.0$, as nebular luminosities depend on \fescion\hs as $\propto (1-f_{\rm esc}^{\rm ion})$ (e.g. Schaerer 2003). The shell model therefore technically only gives rise to Ly$\alpha$ emission while \fescion$\neq 0$ over a finely tuned narrow range of $N_{\rm HI}$ centered on $N_{\rm HI} \sim 10^{17}$ cm$^{-2}$.

Here, we focus on the clumpy ISM models. We have recently constructed a large library of clumpy models (Gronke \& Dijkstra 2016).  In these models, the clumps have HI column densities large enough to make them opaque to LyC photons. However, there exist sightlines that do not penetrate any clumps, and which allow LyC photons to escape. In clumpy models, \fescion \hs is related to the fraction of sightlines from the LyC source(s) which do not intersect any clumps (this corresponds to the `picket fence model' of Heckman et al. 2011). 

The geometry of the clumpy ISM model and its the main parameters are based on that described in Laursen et al. (2013). We refer the interested reader to these papers for a more detailed description on how Laursen et al. (2013) constrain their parameters through observed galaxies. Here, we only present only a brief description of the model. 

In the clumpy ISM model, the multiphase ISM is represented by a large number of neutral, spherical `clumps' which are embedded within a hot gas. The neutral clumps are distributed in a sphere of radius $r_{\rm gal}=5\,$kpc. The clouds themselves have radius $r_{\rm cl}$. The cloud covering factor $f_{\rm cl}$ denotes the total number of clouds from the center of the sphere to its edge, averaged over all sightlines. The content of the cold [hot] clumps [inter-clump medium] is described by $T_{cl},\, n_{\HI, cl}$ [$T_{ICM},\hs n_{\HI, ICM}$] for temperature\footnote{The temperature is defined as $b^2\equiv 2k_{\rm p}T/m_{\rm p}$, where $b^2=v^2_{\rm th}+v^2_{\rm turb}$. Here $v_{\rm th}$ [$v_{\rm turb}$] denotes the thermal [turbulent] velocity of the gas.} and the number density of hydrogen, respectively. The dust optical optical depth through the clouds per path-length given by $\sigma_d Z_{\rm cl}/Z_{\rm sun}n_{\HI}$ where $\sigma_d=1.58\times 10^{-21} \cm^{2}$ \citep{Pei92,Laursen09}, where $Z_{\rm cl}$ denotes the `metallicity' of the cloud (the ICM has metallicity $Z_{ICM}\equiv \zeta_Z Z_{\rm cl}$). Following previous analyses, we assume that there is no further structure to the cold clumps. That is, we do not further split up the neutral clumps into `warm' and `cold' neutral media, as is the case for realistic multiphase gases (e.g. McKee \& Ostriker 1977).

The clumps are outflowing\footnote{Changing the sign of $v(r)$ only `flips' the emerging Ly$\alpha$ spectrum around $x=0$, and leaves our \fescion\hs and \fesclya\hs unaffected.} with a velocity profile 
\begin{equation}
  v(r) = v_{\infty,{\rm cl}}\left\{1 - \left(\frac{r}{r_{\text{min}}}\right)^{1-\beta_{\rm cl}}\right\}^{1/2}
\label{eq:velocity_profile}
\end{equation} for $r>r_{\text{min}} = 1\,{\rm kpc}$ and otherwise zero (Steidel et al. 2010, Laursen et al. 2013). In addition to this, the clouds have a random, isotropic velocity distribution which is Gaussian with a standard deviation $\sigma_{cl}$.

Ly$\alpha$ photons are emitted randomly following an exponential radial volume emissivity profile $\epsilon_{{\rm Ly}\alpha}(r)=\mathcal{N}\exp(-r / H_{\rm em})$ where $\mathcal{N}$ is a normalization constang, and $r$ is the distance to the center of the cloud. The photon is emitted {\it inside} a cloud with probability $P_{\rm cl}$, which would force it to escape from its birthcloud first. The frequency of the photon is drawn from a Gaussian with standard deviation $\sigma_i$. In all our models, we assume the LyC emission traces Ly$\alpha$ emission exactly, including that a fraction $P_{\rm cl}$ is emitted inside a neutral clump. The vast majority of the LyC photons that are emitted inside a cloud do not to escape. 

We thus need $14$ parameters to completely characterize our models\footnote{Note, that the parameters given here differ slightly from what we used in Gronke \& Dijkstra (2014). There, we ignored the filling of the ICM since we were interested in the (enhancement of) the \Lya escape fraction.}. Laursen et al. (2013) discuss plausible ranges for each parameter based on theoretical models, and observations of the ISM in the Milky Way, nearby dwarf galaxies, Ly$\alpha$ emitters (LAEs) and drop-out galaxies out to $z\sim 6$. Our fiducial model adopts the central value of the range quoted in Laursen et al. (2013) as `reasonable', with the exception of the outflow velocity $v_{\infty, {\rm cl}}$ for which Laursen et al. (2013) chose deliberately small values. Values for each parameter are listed in Table~\ref{tab:models} shown in the Appendix. We assembled a library of $2,500$ spectra (using $\sim 10,000$ \textit{escaped} photons each). We drew each parameter uniformly\footnote{Note that $n_{\HI, {\rm ICM}},\,n_{\rm d, ICM},\,T_{\rm ICM},\,T_{\rm cl},\,Z_{\rm cl}$ and $\zeta_Z$ were drawn uniformly in log-space.} from the range indicated in Table~\ref{tab:models}, which is loosely based on the `extreme' range in Laursen et al. (2013). This choice gives us a suite of empirical, simplified  models of the multi-phase ISM that span the wide range of physical conditions encountered in observed galaxies.

\section{Results: Correlation Between \fescion \hs and \fesclya}\label{sec:results}

\begin{figure}
\includegraphics[width=9.0cm,angle=0]{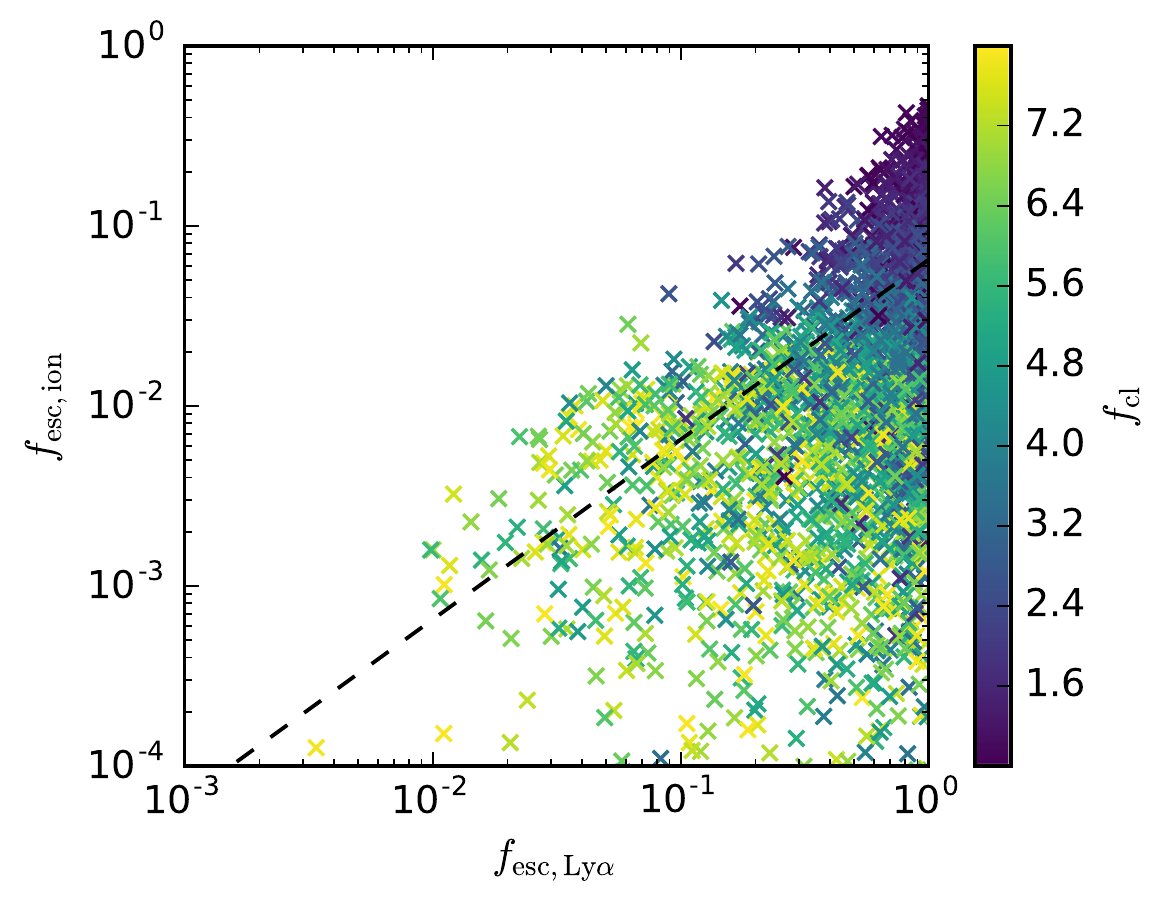}
\vspace{0mm}
\caption[]{The ionizing photon escape fraction, \fescion, as a function of Ly$\alpha$ escape fraction, \fesclya, for a suite of 2500 clumpy ISM models. Each {\it cross} represents the angle-averaged escape fraction for a complete Ly$\alpha$ Monte-Carlo radiative transfer calculation for one particular parametrization of the clumpy ISM model. The {\it color} of the {\it crosses} denote the cloud covering factor $f_{\rm cl}$. This plot shows that there is a correlation between the two parameters: galaxies with low \fesclya \hs have a low \fescion, while galaxies with high \fesclya \hs show a large spread in \fescion, driven strongly by $f_{\rm cl}$. }
\label{fig:1}
\end{figure}

Figure~\ref{fig:1} shows \fescion \hs as a function of \fesclya. Each cross represents a Monte-Carlo radiative transfer simulation for one random realization of a clumpy ISM model. The color of the cross denotes $f_{\rm cl}$. 

There are several take-away points from this plot.
\begin{enumerate}[leftmargin=*]
\item The $2500$ models give rise to significant variation in \fesclya\hs (which spans $\sim$ 3 orders of magnitude) and \fescion\hs (which spans $\sim 4$ orders of magnitude). Models that give rise to \fesclya $\gsim 0.1-0.2$ would correspond to galaxies with relatively `strong' Ly$\alpha$ emission, such as Ly$\alpha$ emitters. Our models therefore give rise to a population of `Ly$\alpha$ emitters' and weaker Ly$\alpha$ sources such as drop-out galaxies with weak Ly$\alpha$ emission. Our results also indicate that for at fixed \fesclya, the dispersion in \fescion\hs can be large.

\item When \fesclya\hs is small, then \fescion\hs is small. Ly$\alpha$ photons are destroyed most efficiently when they encounter, and scatter in, many different clumps. The number of scattering events (or `cloud interactions') scales as $\mathcal{N}_{\rm cl}\propto f^2_{\rm cl}$ ($f_{\rm cl} \gg 1$, Hansen \& Oh 2006). If $f_{\rm cl}$ is larger, then the Poission probability of having clear sightlines becomes exponentially smaller: the Poisson probability that a sightline from $r=0$ intersects zero clumps equals\footnote{The expression for $P(N_{\rm clump}=0|f_{\rm cl})$ which properly includes the Ly$\alpha$ emissivity profile $\epsilon_{{\rm Ly}\alpha}(r)$ must take into account that the probability $P(N_{\rm clump}=0|f_{\rm cl},r)$ depends on emission direction for $r \neq 0$. This makes the expression for $P(N_{\rm clump}=0|f_{\rm cl})$ a bit more complicated but preserves the exponential dependence on $f_{\rm cl}$. } $P(N_{\rm clump}=0|f_{\rm cl},r=0)=(1-P_{\rm cloud})\exp(-f_{\rm cl})$, where $1-P_{\rm cloud}$ denotes the probability that the LyC photon was {\it not} emitted inside a cloud. 

This result may make it difficult to explain inferred \fescion$\sim 0.1-0.2$ for a small subset LBGs (e.g. Iwata et al. 2009, Micheva et al. 2016). This apparent discrepancy can be alleviated in five ways: ({\it i}) resonant scattering of Ly$\alpha$ off residual HI gas in the diffuse IGM can suppress the observed Ly$\alpha$ flux by an additional factor of $1.5-2.0$ depending on redshift (e.g. Laursen et al. 2011), which should be applied to our predicted \fesclya\hs when comparing to observations;  ({\it ii}) the fraction of LBGs with claimed LyC detections is very small, which suggests this population is rare, and not captured by our analysis in spite of our coverage of a broad range of ISM physical conditions; ({\it iii}) While LBGs generally have smaller \fesclya\hs than LAEs, \fesclya\hs appears correlated with the Ly$\alpha$ EW (e.g. Trainor et al. 2015, Micheva et al. 2016), which itself scales as EW$\propto (1-$\fescion$)$\fesclya\hs(see \S~\ref{sec:lyaprod}). This suggests that those LBGs that show LyC leakage, may in fact have larger \fesclya\hs than the LBG population as a whole; ({\it iv}) For very large \fescion\hs, the production rate of Ly$\alpha$ decreases, which mimicks a low \fesclya\hs (see \S~\ref{sec:lyaprod} for more discussion on this); ({\it v}) Each cross in Figure~\ref{fig:1} represents an {\it angle-average} of the escape fractions for each of the 2500 models. The `apparent' \fescion\hs can be larger along sightlines which do not intersect any clumps. The Ly$\alpha$ escape fraction can also be enhanced for these same sightlines, though scattering of Ly$\alpha$ photons suppresses the angular variation of \fesclya\hs (see Gronke \& Dijkstra 2014). The angular variation of both escape fraction can be represented by replacing each cross in Figure~\ref{fig:1} with a distribution which is elongated along the \fescion-direction, which may help explain that objects exist for which \fescion\hs is high, while \fesclya\hs is low.

\item The dispersion in \fescion\hs at fixed \fesclya \hs increases with \fesclya. In other words, as we increase \fesclya\hs the probability of having a large \fescion\hs increases. There are a number of ways to boost \fesclya. These include reducing the dust content of the neutral clumps, increasing the outflow velocity, and reducing $f_{\rm cl}$. As mentioned above, reducing $f_{\rm cl}$ enhance the Poisson probability that there exist sightlines that do not intersect any clumps, which increases \fescion. \citet{Matthee16b} recently found \fescion$\gsim 60\%$ for 8 H$\alpha$ emitters (HAEs) out of a sample of 191. Two of these LyC emitting HAEs have a high \fesclya\hs (Matthee et al. 2016a). For the remaining 6 the Ly$\alpha$ is not good enough (yet) to constrain \fesclya.

\item The color coding of the {\it crosses} in Figure~\ref{fig:1} show clearly that high \fescion\hs are those with low $f_{\rm cl}$. This again reflects that a lower average number density of clouds from the center to the edge of the `galaxy' boosts the Poisson probability for having clean sightlines. 

\end{enumerate}

The strong $f_{\rm cl}$-dependence of \fescion\hs is easily understood from analytic arguments. The simulations indicate that this result is not significantly affected by varying the other parameters. The {\it black dashed line} shows the best linear fit through the collection of data points. We stress that the purpose of this line is to illustrate that \fescion\hs and \fesclya\hs are correlated. The exact `best-fit' correlation depends on how the 14 model-parameters were sampled: different PDFs for the model parameters would likely yield a different best-fit correlation. This may help explain that our correlation differs quantitatively from that found by Yajima et al. (2014), who found few objects with high \fesclya\hs and low \fescion. This difference may also reflect that ({\it i}) the simulations do not resolve the multi-phase interstellar medium, and may therefore not properly capture that Ly$\alpha$ photons avoid destruction by dust by scattering off the surface of dense, neutral clumps which contain most of the dust, ({\it ii}) that our model artificially enhances this surface scattering effect, by representing the multi-phase ISM as a two-phase medium. We stress that the purpose of our calculations was not to derive the correct correlation, which would be overambitious, but rather to show that for reasonable parameters for the multiphase ISM, a correlation exists. Finally, it is worth mentioning that the fact that both \fescion\hs and \fesclya\hs are affected most strongly by $f_{\rm cl}$ implies that the precise structure of the clumps (i.e. the presence of a `cold neutral medium' inside the clumps), would introduce changes that are subdominant to those introduced by $f_{\rm cl}$.

\section{Discussion}\label{sec:discuss}

\subsection{The $M_{\rm UV}$-dependence of \fesclya}\label{sec:muvfesc}

The `Ly$\alpha$ fraction' denotes the fraction of galaxies that have a Ly$\alpha$ emission line stronger than some threshold equivalent width. Observations indicate that the Ly$\alpha$ fraction increases with $M_{\rm UV}$ (e.g. Stark et al. 2010, Pentericci et al. 2011, Caruana et al. 2012, Ono et al. 2012, Schenker et al. 2012, Pentericci et al. 2014, Caruana et al. 2014). Gronke et al. (2015b) combined observations of the UV-LF with current constraints on the UV-dependence of the Ly$\alpha$ fraction, and predicted that Ly$\alpha$-LFs should be have steeper faint ends than the UV-LFs. Specifically, if we denote the faint-end slope of the Ly$\alpha$ LF with $\alpha_{{\rm Ly}\alpha}$, then $\alpha_{{\rm Ly}\alpha}=\alpha_{\rm UV}-x$ where $x\sim 0.2-0.4$ (see Fig~2 of Gronke et al. 2015b). Recent measurements of faint end slope of Ly$\alpha$ emitter luminosity functions at $z=5.7$ indicate that $\alpha_{{\rm Ly}\alpha}\sim -2.2\pm 0.2$ (Dressler et al. 2015), and that $\alpha_{{\rm Ly}\alpha}\sim -1.75\pm 0.1$ at $z\sim 2$ (Konno et al. 2016). These measurements agree well\footnote{Gronke et al. (2015b) only predicted Ly$\alpha$ LFs at $z\geq 3$. We extrapolated their predictions for $\alpha_{{\rm Ly}\alpha}$ to $z\sim 2$. This same extrapolation would translate to a faint-end slope of the UV-luminosity function at $z\sim 2.3$ that is $\alpha_{\rm UV} \sim -1.5$, which agrees with recent determinations (see Fig~10 of Parsa et al. 2016), though not all (see e.g. Reddy \& Steidel 2009, who found a steeper $\alpha_{\rm UV}=-1.73\pm 0.07$).} with prediction using Ly$\alpha$ fraction constraints, and provides independent confirmation that more Ly$\alpha$ radiation emerges per `unit' UV-flux density towards lower UV luminosities.

 This enhanced emergence of Ly$\alpha$ flux from UV-faint galaxies implies that ({\it i}) the Ly$\alpha$ production rate increases, and/or ({\it ii}) \fesclya\hs increases towards lower UV-luminosities. 
 Recent work has shown that at $z\sim 4$ the ionizing photon production efficiency, $\xi_{\rm ion}$ (Robertson et al. 2013), appears to be independent of $M_{\rm UV}$ in the range $-21<M_{\rm UV} <-19$ at $z\sim 4-5$ (see Fig~1 of Bouwens et al. 2016, which also shows that there is still a large scatter). The Ly$\alpha$ production efficiency should then also not depend on $M_{\rm UV}$, as Ly$\alpha$ production is directly tied to ionizing photon production. In contrast, over this same range in $M_{\rm UV}$, the Ly$\alpha$ fraction rises rapidly (see Fig~13 of Stark et al. 2010). This suggests that the enhanced visibility of Ly$\alpha$ flux is mostly driven by an enhanced escape fraction, and provides the basis for our statement that there is observational support that \fesclya\hs increases towards lower UV-luminosities (or towards higher $M_{\rm UV}$). Trainor et al. (2015) note that in LAEs with H$\alpha$ detections, the inferred \fesclya\hs correlates significantly with Ly$\alpha$ EW, which provides independent confirmation that Ly$\alpha$ EW is an indicator of Ly$\alpha$ escape.

\citet{Oya16} recently found that the Ly$\alpha$-EW PDF, and therefore the Ly$\alpha$ fraction, depends on stellar mass, $M_*$. This supports that \fesclya\hs increases towards lower $M_*$. Our finding of a correlation between \fesclya\hs and \fescion\hs then implies that \fescion\hs also increases towards lower $M_*$. Faisst (2016) independently came to this conclusion by combining the observed correlation between \fescion\hs and the [O III]$\lambda$5007/[O II]$\lambda$3727 line ratio, and the (anti-)correlation of this line ratio with $M_*$ inferred from local high-z analogues. We stress that we focus on the $M_{\rm UV}$-dependence of \fescion\hs because this allows us to directly connect our results to the UV-LF of continuum selected galaxies, which is routinely used to quantify the LyC volume emissivity of galaxies during cosmic reionization.

\subsection{Implications for Reionization}\label{sec:reionization}

One of the main open questions in reionization is whether galaxies provided enough photons to reionize the Universe, and if so, which galaxies provided the dominant contribution to the ionizing background that drove reionization. These questions are commonly addressed by extrapolating the faint-end of the (non-ionizing) UV-LF of drop-out galaxies to some minimum UV luminosity (corresponding to a maximum $M^{\rm lim}_{\rm UV}$), and then see whether theses galaxies provided enough photons to either reionize the Universe, or to keep it ionized \citep[e.g.][]{Wilkins11,Shull12,Kuhlen,Fink,Robertson}. This approach introduces two parameters related to the UV-LF: ({\it i}) its faint end slope ($\alpha_{\rm UV}$), and ({\it ii}) its minimum cut-off luminosity ($M^{\rm lim}_{\rm UV}$). For a fixed set of parameters $(\alpha_{\rm UV},M^{\rm lim}_{\rm UV})$, the question whether galaxies reionized the Universe then translates to a constraint on \fescion. This constraint on \fescion\hs represents a (weighted) average over the full population of UV emitting galaxies. There have been numerous theoretical efforts to model the faint end slope of the UV-LF and where it may flatten (e.g. Jaacks et al. 2013, Mason et al. 2015, O'Shea et al. 2015, Liu et al. 2016).

\begin{figure}
\vspace{-10mm}
\includegraphics[width=10.0cm,angle=0]{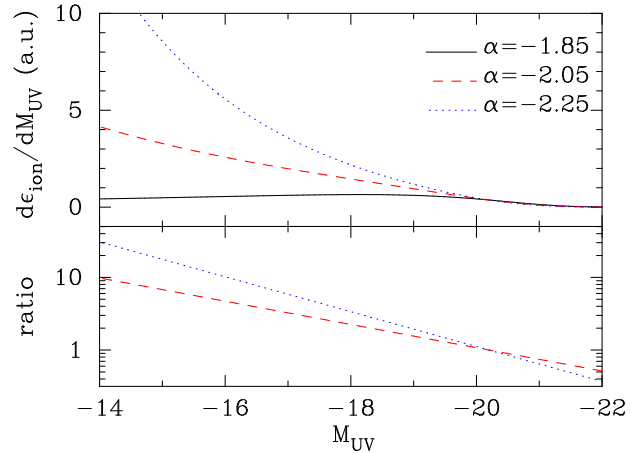}
\vspace{-5mm}
\caption[]{An increase in \fescion\hs towards lower UV luminosities gives rise to a steepening of the LyC luminosity function (LF), which can be mimicked with a steeper UV LF and a constant \fescion. The {\it top panel} of this Figure shows the relative contribution $d\epsilon_{\rm ion}/dM_{\rm UV}$ (in arbitrary units) to the ionizing volume emissivity $\epsilon_{\rm ion}$ at $z=6$ by galaxies with $M_{\rm UV} \pm dM_{\rm UV}/2$ for the measured $\alpha_{\rm UV}=-1.85$ ({\it black solid line}), and steeper LFs with $\alpha_{\rm UV}=-2.25$ ({\it blue dotted line}) and $\alpha_{\rm UV}=-2.05$ ({\it red dashed line}). While we cannot predict (yet) which $\alpha_{\rm UV}$ mimicks the true $M_{\rm UV}$-dependence of \fescion, this plot visually illustrates the enhanced contribution of UV-faint galaxies to cosmic reionization. The {\it lower panel} shows the ratio between the models in the {\it top panel} and the fiducial model (shown as the {\it black solid line}).}
\label{fig:3}
\end{figure} 

If Ly$\alpha$ and LyC escape are correlated, then we also expect \fescion\hs to increase towards lower UV luminosities. Just like the case for Ly$\alpha$, if we were to plot the LyC luminosity function (i.e. the number density of galaxies as a function of LyC luminosity), it would be steeper than the UV luminosity function. This steepening can be mimicked by a model in which \fescion\hs does not depend on $M_{\rm UV}$, and in which the faint-end slope of the UV-luminosity function is made steeper. Figure~\ref{fig:3} visually illustrates the impact of this steepening, and the {\it top panel} shows the relative contribution $d\epsilon_{\rm ion}/dM_{\rm UV}$ to the ionizing volume emissivity $\epsilon_{\rm ion}$ by galaxies in the range $M_{\rm UV} \pm dM_{\rm UV}/2$. The {\it black solid line} shows $d\epsilon_{\rm ion}/dM_{\rm UV}$ for the `standard' Schechter function parameters at $z=6$, $(\alpha_{\rm UV}, M_*)=(-1.85,-20.2)$ (using the fitting formula from Bouwens et al. 2015). For the {\it blue dotted line} [{\it red dashed line}] we increased $\alpha_{\rm UV} \rightarrow -2.25$, which represents the steepening relevant for the Ly$\alpha$ LF [$\alpha_{\rm UV} \rightarrow -2.05$, which represents an intermediate case]. While we do not know which $\alpha_{\rm UV}$ mimicks the correct $M_{\rm UV}$ dependence of \fescion, it does illustrate the possible enhanced contribution of UV-faint galaxies to cosmic reionization\footnote{Decreasing $\alpha$ reduces the contribution of UV-bright galaxies, i.e. $M_{\rm UV} < M_*$, to $d\epsilon_{\rm ion}/dM_{\rm UV}$ because reducing $\alpha$ also affects the bright end of the luminosity function.}. The enhancement is illustrated in the {\it lower panel} of Figure~\ref{fig:3} which shows the ratio of the models shown in the {\it top panel}. This plot shows that in a model with $\alpha_{\rm UV}=-2.25$ galaxies with $M_{\rm UV} \sim -16$ contribute $\gsim 10$ times more to the total ionizing photon production rate than when $\alpha=-1.85$. This model extends down to $M_{\rm UV}=-14$, which corresponds (roughly) to the limit to which the UV-LF has been constraint to be a power-law \citep[see e.g][also see O'Shea et al. 2015 and Liu et al. 2016 for theoretical arguments why the UV-LF may flatten at $M_{\rm UV} \gsim -14$]{Alavi14,Parsa16,Livermore16}.

\begin{figure*}
\centering
\includegraphics[width=16.0cm,angle=0]{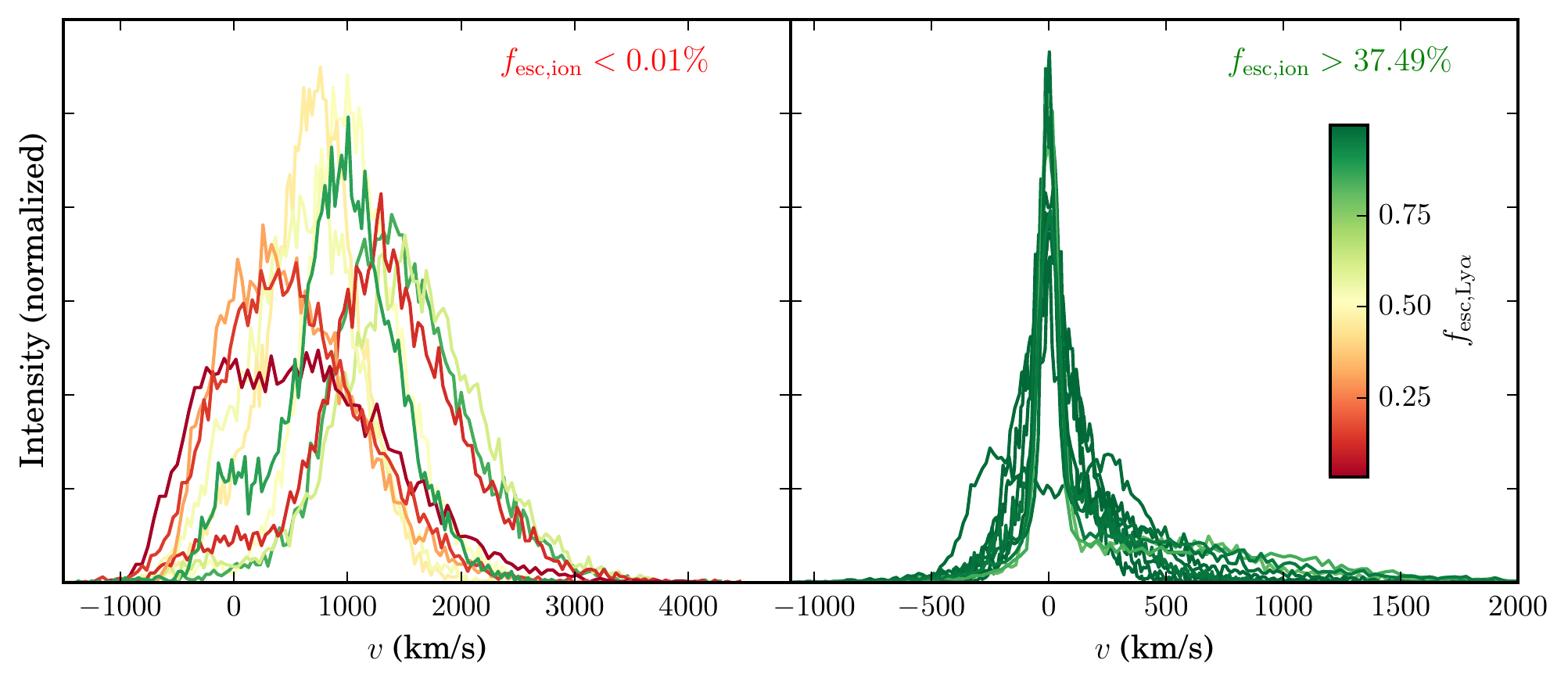}
\vspace{0mm}
\caption[]{{\it Left panel}: Ly$\alpha$ spectra emerging from 25 models with the lowest \fescion$<10^{-4}$. {\it Right panel}:  Ly$\alpha$ spectra emerging from the $25$ models with the highest \fescion$>0.37$. This Figure shows that for models with low \fescion\hs Ly$\alpha$ spectra are redshifted, asymmetric, and broad. The width and velocity off-set of the models with the {\it lowest} \fescion\hs are larger than what has been observed, which is likely because of the simplified representation of the multiphase ISM (see text). Although models with high \fescion\hs exhibit a variety of spectral line shapes, their spectra are generally narrower and more symmetric.}
\label{fig:spec}
\end{figure*}

Finally, Ly$\alpha$ escape in clumpy ISM models -- which were introduced to reflect the multi-phase nature of the ISM -- is most strongly regulated by covering factor, and to a lesser extent by other parameters such as the dust content (Gronke \& Dijkstra 2016). Star forming galaxies are known to become bluer towards higher redshift, which is taken as evidence that star forming galaxies get less dusty towards higher redshifts \citep[e.g.][]{Finkcolors,Bouwenscolor}. Our conclusions would break down if Ly$\alpha$ escape were driven {\it entirely} by the changing dust content of an otherwise identical scattering medium. In this case however, we would expect both the width and (possibly) velocity shift of the Ly$\alpha$ line to {\it increase} with $M_{\rm UV}$, because Ly$\alpha$ scattering causes photons to diffuse in frequency space, and to broaden the Ly$\alpha$ spectral line shape. If only dust were regulating Ly$\alpha$ escape, then dust would suppress this frequency diffusion and cause lines to be narrower (see e.g. Fig~8 of Laursen et al. 2009). The evolution in Ly$\alpha$ line width and shift predicted by the `pure dust' scenario is not consistent with observations which indicate that Ly$\alpha$ spectra of Ly$\alpha$ emitting galaxies, if anything, tend to get narrower: Konno et al. (2016) have shown that shell-model fits to observed Ly$\alpha$ line profiles favor increasingly low HI column densities towards higher $z$ for otherwise identical shell model parameters. The reduced HI column density introduces less frequency diffusion, and makes Ly$\alpha$ line profiles narrower\footnote{Also, the velocity off-set of the peak flux density of the Ly$\alpha$ spectral line shape decreases towards UV-fainter galaxies \citep[e.g.][]{Erb14,Song14}.}.  In addition, there is observational support that the covering factor of low-ionization metal lines decreases with $z$ (e.g. Jones et al. 2013). If these metals trace cold, neutral gas, then this supports the notion that Ly$\alpha$ escape increases towards higher redshift (at least partly) because of the evolution in the covering factor of neutral gas.

\subsection{Connection \fescion\hs to the Ly$\alpha$ Spectrum}\label{sec:spectrum}

Figure~\ref{fig:1} showed that \fescion\hs depends sensitively on $f_{\rm cl}$, which was due to the exponential dependence on $f_{\rm cl}$ of the Poisson probability of having sightlines with no clumps. The parameter $f_{\rm cl}$ is known to play a key role in Ly$\alpha$ transfer through clumpy media (Hansen \& Oh 2006). We have demonstrated that $f_{\rm cl}$ is one the 14 parameters of the clumpy models that most strongly affects the emerging Ly$\alpha$ spectrum (Gronke \& Dijkstra 2016). This implies immediately that \fescion\hs should be closely correlated with spectral features of the Ly$\alpha$ line. 

Figure~\ref{fig:spec} compares Ly$\alpha$ spectra for 25 models with the {\it highest} \fescion$>0.37$ ({\it right panel}) to 25 models with the {\it lowest} \fescion$<10^{-4}$ ({\it left panel}).  These two panels illustrate clearly that a high \fescion\hs corresponds to having narrower, more symmetric Ly$\alpha$ lines. Models that have the highest \fescion\hs show a variety in their spectra. We caution that the width and velocity off-set of the models with the {\it lowest} \fescion\hs are larger than what has been observed. This is likely an artefact of the models: models with the lowest \fescion\hs have the highest $f_{\rm cl} \sim 8$. Ly$\alpha$ photons typically scatter off $\sim f^2_{\rm CL}$ separate clouds before escaping (e.g. Hansen \& Oh 2006), and each `cloud-interaction' can impart of noticeable Doppler boost on the Ly$\alpha$ photon, which broadens the Ly$\alpha$ spectral line.

The connection between the Ly$\alpha$ spectral shape and ionizing photon escape was pointed out previously by Behrens et al. (2014, in the context of modified shell models) and Verhamme et al. (2015, in the context of shell models). In these models, LyC escape translated to ({\it i}) significant Ly$\alpha$ flux at systemic velocity and/or ({\it ii}) a small peak separation ($\Delta v \lsim 300$ km s$^{-1}$). In multiphase models, it is not possible to point out features in the spectrum that guarantee a LyC detection, partly because of the larger variety in the spectra associated with models that have larger \fescion. In addition, in clumpy models Ly$\alpha$ photons can escape after scattering off a single gas cloud, and close to the frequency at which they were initially emitted (also see Hansen\& Oh 2006, Laursen et al. 2013). Moreover, while narrow Ly$\alpha$ lines that are symmetric around the systemic velocity of the host galaxy translate to a higher probability of being a LyC emitting galaxy, LyC escape is highly anisotropic (Ly$\alpha$ escape less so, see Gronke \& Dijkstra 2014), which further complicates making robust predictions for whether we can observe LyC flux from a galaxy or not. However, anisotropic escape of LyC photons similarly affects other promising LyC-leakage indicators\footnote{On the other hand, LyC escape enhances the ionizing radiation field in close proximity to star forming galaxies, which can increase the surface brightness in fluorescent Ly$\alpha$ and H$\alpha$ emission (see Mas-Ribas \& Dijkstra 2016).} such as the [O III]$\lambda$5007/[O II]$\lambda$3727 line ratio (Jaskot \& Oey 2013, Nakajima \& Ouchi 2014). 

The low-redshift `Lyman Break Analogue' (LBA, Heckman et al. 2011, Borthakur et al. 2015) and `green pea galaxy' (Henry et al. 2015, Yang et a. 2016, Izotov et al. 2016) with reported detections of LyC escape, had unusual Ly$\alpha$ spectra in the sense that the spectra contained significant flux blueward of Ly$\alpha$ resonance. These spectra were different than those shown in Figure~\ref{fig:spec} in that they had deep `absorption' troughs separating the blue and red peaks, which are absent from the spectra in Figure~\ref{fig:spec}. The absence of these absorption troughs in the theoretical spectra may reflect the lack trace amounts of residual HI (possibly in the CGM) at systemic velocity (see Gronke \& Dijkstra 2016). In any case, the presence of flux blueward of the Ly$\alpha$ resonance indicates that the lines are more symmetric around the Ly$\alpha$ resonance than is common.

\subsection{Suppressed Ly$\alpha$ Production for large \fescion}\label{sec:lyaprod}

The Ly$\alpha$ production rate scales as $\propto (1-$\fescion$)$. The total Ly$\alpha$ flux that we receive from a distant galaxy, as well as the equivalent width (EW) of the line, both scale as $\propto (1-$\fescion$)$\fesclya.

\begin{figure}
\includegraphics[width=9.0cm,angle=0]{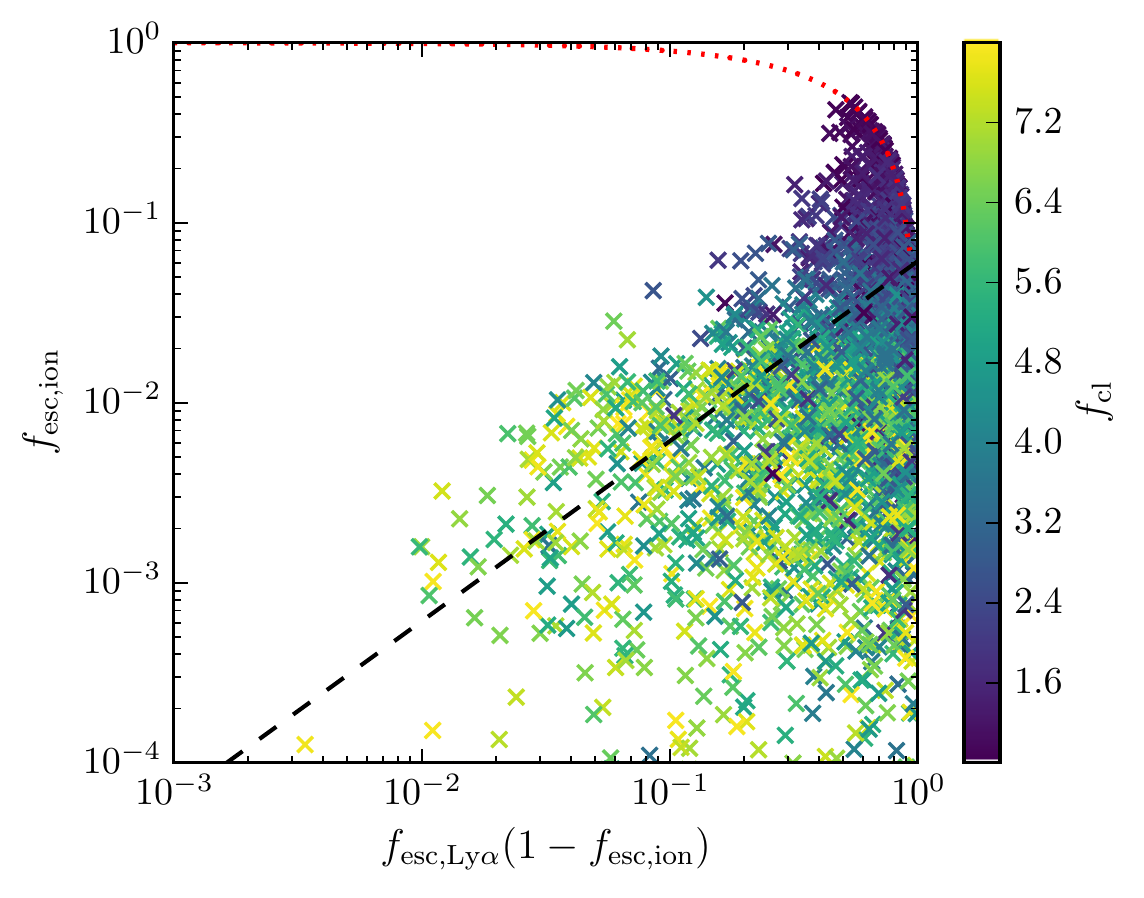}
\vspace{0mm}
\caption[]{This plot shows \fescion\hs as a function of the `apparent' Ly$\alpha$ escape fraction, \fesclya$(1-$\fescion$)$, which reflects that the production rate of Ly$\alpha$ photons scales as $\propto (1-$\fescion$)$. The {\it red-dotted line} shows the maximum apparent escape fraction $(1-$\fescion$)$. Large \fescion\hs thus also suppresses the observed Ly$\alpha$ flux, mimicking a reduction in \fesclya. This Figure illustrates that there exists a maximum average \fesclya$(1-$\fescion$)$ at some \fescion$\equiv$\fescionmx$\sim 0.1-0.5$ (see text).}
\label{fig:5}
\end{figure}

Figure~\ref{fig:5} shows $\log$\fescion \hs as a function of $\log$[\fesclya$(1-$\fescion$)]$. The turnover at high-\fescion\hs and high-\fesclya\hs reflects that the quantity \fesclya$(1-$\fescion$)$ cannot exceed $(1-$\fescion$)$ (which is indicated as the {\it red dotted line}). At fixed \fescion\hs there exists a distribution of  \fesclya$(1-$\fescion$)$, which reflects the dispersion in \fesclya. The average of this distribution peaks at some maximum \fescionmx\hs(also see Dijkstra et al. 2014). The value of \fescionmx\hs is model-dependent, and even in the context of our model it depends on how we sampled our 14 parameters. It nevertheless seems reasonable to assume that \fescionmx$\sim 0.1-0.5$. For large \fescion$>$\fescionmx\hs the Ly$\alpha$ luminosity drops again, which mimicks a reduction in \fesclya. We expect this `apparent' reduction in the escape fraction to translate to a reduction in the Ly$\alpha$ fraction and/or a flattening of the Ly$\alpha$ luminosity function at low Ly$\alpha$ luminosities\footnote{Although a flattening has possibly been detected at $z\sim 3$ by Rauch et al. (2008) at $L_{\alpha} \sim 10^{41}$ erg s$^{-1}$, which would probe galaxies with $M_{\rm UV} \sim -15 \pm 1$ (see Gronke et al. 2015 for a discussion).}. No evidence for either this drop or this flattening exists at $z\sim 6$ in current data (though it may be present at $z~6.5$, see Fig~7 of Matthee et al. 2015), which implies that this effect is not important in current observations at $z\sim 6$. More quantitatively, Dressler et al. (2015) infer a steep faint end slope of the LAE LF down to $L_{\alpha} < 10^{42}$ erg s$^{-1}$. Gronke et al. (2015b) show that Ly$\alpha$ luminosity of $L_{\alpha} \sim 10^{42}$ erg s$^{-1}$ probes galaxies with $M_{\rm UV} \sim -18 \pm 1$ (see their Fig~3). This therefore implies hat \fescion$<$\fescionmx, and that therefore this effect is not important, down to $M_{\rm UV} \sim -18 \pm 1$. 

At $z>6$ there is observational evidence for a reduction in the Ly$\alpha$ flux from star forming galaxies compared to expectations based on extrapolations from lower redshift observations (see e.g. Dijkstra 2014 and references therein). There are indications that this reduction is more severe for UV-faint galaxies (e.g. Ono et al. 2012, Pentericci et al. 2014), which is commonly interpreted as a signature of inhomogeneous reionization, but might also reflect that \fescion$\rightarrow$\fescionmx\hs in UV-faint galaxies at $z\sim 7$ (also see Dijkstra et al. 2014). It is theoretically possible to distinguish between these two scenarios: ({\it i}) reionization leaves a unique signature on the angular clustering of Ly$\alpha$ emitters (McQuinn et al. 2007, Mesinger \& Furlanetto 2008, Jensen et al. 2013, Sobacchi \& Mesinger 2015), and which can be measured with Subaru's Hyper-Suprime Cam\footnote{\url{http://www.naoj.org/Projects/HSC/}} (see e.g. Jensen et al. 2014, Sobacchi \& Mesinger 2015), ({\it ii}) if Ly$\alpha$ disappears as a result of \fescion\hs becoming large, then we should see a similar decrease in the line strength of other non-resonant nebular lines such as H$\alpha$ (and H$\beta$), something that can be tested with the James Webb Space Telescope\footnote{\url{http://www.jwst.nasa.gov/}} \citep{JWST}.

Redshift $z\sim 6$ is particularly interesting as reionization likely had little impact on the observed Ly$\alpha$ flux from galaxies. Should future data reveal a flattening in the Ly$\alpha$ LF at low Ly$\alpha$ luminosities and/or a reduction in the Ly$\alpha$ fraction at lower UV-luminosities, then this may provide a valuable constraint on \fescion\hs at this redshift. In addition, understanding whether \fescion\hs introduces a flattening in the Ly$\alpha$ LF at low Ly$\alpha$ luminosities and/or a drop in the Ly$\alpha$ fraction at faint UV luminosities, would help us better constrain the role that reionization plays in suppressing the Ly$\alpha$ flux from galaxies at $z>6$.

\subsection{Impact of Delayed Ly$\alpha$ Escape due to Trapping}
The escape fraction of LyC photons from a galaxy can vary significantly on time-scales of $\sim 10$ Myr \citep[][]{Kimm14,Ma15}, which corresponds approximately to the life-time of massive stars. Trapping of Ly$\alpha$ photons by HI gas can introduce a lag in the escape of Ly$\alpha$ and LyC photons (Yajima \& Li 2014): Ly$\alpha$ photons scatter inside HI gas when \fescion$ \ll 1$, but are `released' efficiently when low-column channels temporarily open-up, which allow LyC photons to escape. Time-variations in \fescion\hs and delayed escape of Ly$\alpha$ has only a minor, positive, impact on our results by slightly more tightly coupling \fesclya\hs and \fescion, as we explain below.

Trapping of Ly$\alpha$ photons is limited to time-scales $t_{\rm trap} \ll 10$ Myr, as the typical Ly$\alpha$ trapping time equals $t_{\rm trap}=|x_{\rm p}|t_{\rm cross}$, where $t_{\rm cross}\equiv R/c$ denotes the time it takes radiation to escape in the absence of scattering, and $|x_{\rm p}|\approx 12(N_{\rm HI}/10^{20}\hs{\rm cm^{-2}})^{1/3}(T/10^4\hs{\rm K})^{1/6}$ for a static, uniform, spherical gas cloud with an HI column density $N_{\rm HI}$ and temperature $T$ (Adams, 1975). In reality, this estimate provides a strict upper limit to the delay time: velocity gradients, density inhomogeneities reduce $t_{\rm trap}$ (Bonilha et al. 1979, Dijkstra \& Loeb 2008). Laursen et al. (2013) evaluated that the typical trapping time for Ly$\alpha$ in clumpy media considered here to be $t_{\rm trap} \sim 2 \times 10^4$ yr. Trapping of Ly$\alpha$ photons is therefore unlikely to introduce a lag between the escape of Ly$\alpha$ and LyC photons at a level where it has observable consequences. Moreover, if anything, this effect would serve to more tightly couple Ly$\alpha$ and LyC escape, as LyC escape would be accompanied with the escape of Ly$\alpha$ photons that were trapped inside the HI gas. \\

\section{Conclusions}
\label{sec:conc}

The escape fraction of ionizing photons, \fescion, represents one of the great unknowns in our understanding of cosmic reionization. Observational constraints on \fescion\hs are still weak, and theoretical predictions remain incomplete owing to the challenging nature of the calculations. We have computed the correlation between the escape fractions of Ly$\alpha$ (\fesclya) and ionizing (LyC) radiation (\fescion) by performing Monte-Carlo simulations of Ly$\alpha$  radiative transfer through a suite of $2500$ models of dusty, clumpy interstellar media. This represents a `top-down' (empirical) approach to modelling LyC and Ly$\alpha$ transfer through realistic, multiphase interstellar media, and complements the previous `bottom-up' (ab initio) approach by Yajima et al. (2014), who used hydrodynamical simulations to generate models of the ISM. Our main results are:

\begin{itemize}[leftmargin=0pt,itemindent=20pt]
\item We find that \fescion \hs and \fesclya \hs are correlated. The dispersion in \fescion \hs at fixed \fesclya \hs increases towards larger \fesclya: galaxies with low \fesclya \hs have a low \fescion, while galaxies with high \fesclya \hs show a large spread in \fescion\hs (see Fig~\ref{fig:1}). The dispersion in \fescion\hs is driven by the dispersion in $f_{\rm cl}$, which measures the cloud covering factor. Our results agree qualitatively with those obtained by Yajima et al (2014, who also found a positive correlation), but quantitatively some differences remain, which reflects that neither approach has converged yet (see the discussion in \S~\ref{sec:model}). 

While predictions of both \fescion\hs and \fesclya\hs are still highly uncertain, the existence of a correlation between the two quantities can be predicted more robustly, which is underlined by the fact that two different, independent approaches confirm the existence of this correlation. The \fesclya-\fescion\hs correlation reflects that the escape of ionizing photons requires that sightlines exist which contain low column densities of atomic hydrogen, i.e. $N_{\rm HI} \lsim 1/\sigma_{\rm ion} \approx 10^{17}$ cm$^{-2}$. These same-low column density paths provide escape routes for Ly$\alpha$ photons (also see Behrens et al. 2014, Verhamme et al. 2015). At a deeper level, the escape of Ly$\alpha$ is facilitated by outflows, which may also create low column density holes out of galaxies, which in turn permit LyC photons to escape.

\item We argued that the positive correlation between \fescion\hs and \fesclya\hs is directly relevant for studies of cosmic reionization, as there is increasing observational support from both continuum and line selected galaxies that Ly$\alpha$ escapes more easily from UV-faint galaxies (see \S~\ref{sec:muvfesc}). The correlation between \fescion\hs and \fesclya \hs then implies that ionizing photons also escape more easily from UV-faint galaxies at this redshift. This implies that UV-faint galaxies contribute more to the volume emissivity of ionizing photons than implied by the faint-end slope of the UV-luminosity function (\S~\ref{sec:reionization}). These conclusions may be invalidated if the escape of Ly$\alpha$ is regulated {\it purely} by dust. However, we argued in \S~\ref{sec:reionization} that observations do not support this picture. 

\item Because the `apparent' Ly$\alpha$ escape fraction, \fesclya$(1-$\fescion$)$, reaches a maximum value for \fescion$=$\fescionmx$\sim 0.1-0.5$ (see \S~\ref{sec:lyaprod}), we expect a drop in the Ly$\alpha$ fraction at lower UV-luminosities and/or a flattening of the Ly$\alpha$ LF at lower Ly$\alpha$ luminosities, if \fescion\hs continues to rise monotonically. This has not been observed yet at $z\sim 6$ (but possibly at $z\sim 6.5$, see Matthee et al. 2015), which implies that \fescion$<$\fescionmx\hs in galaxies with $M_{\rm UV} \sim -18 \pm 1$. The observed reduction in Ly$\alpha$ flux from galaxies at $z>6$ may be partly due to \fescion\hs approaching \fescionmx\hs (also see Dijkstra et al. 2014). LAE clustering measurements and observations of Balmer lines can help determine the role of \fescion\hs in the disappearance of Ly$\alpha$ emission from galaxies at z$\gsim 6$ (see \S~\ref{sec:lyaprod}).

\item Figure~\ref{fig:1} also shows that the ionizing escape fraction is strongly affected by the cloud covering factor, $f_{\rm cl}$. As a result, \fescion\hs is closely connected to the observed Ly$\alpha$ spectral line shape (see \S~\ref{sec:spectrum}) with LyC emitting objects typically having narrower, more symmetric Ly$\alpha$ lines (Fig~\ref{fig:spec}, also see Erb et al. 2014). In multiphase models, LyC emitting object exhibit a wide range of spectral line profiles, and it is not possible to identify spectral features that `guarantee' a LyC detection. 
\end{itemize}

Ly$\alpha$ emitting galaxies are valuable for constraining the ionization state of the intergalactic medium (see e.g. Dijkstra 2014, and references therein). Our work implies that these galaxies also provide unique insights into the nature of the sources that reionized the Universe, in spite of the fact that modeling interstellar Ly$\alpha$ radiative transfer remains highly challenging. We emphasize that our results differ from previous works which estimated the contribution of LAEs to cosmic reionization (see e.g. Yajima et al. 2014): LAEs represent a subset of galaxies within a limited range of $M_{\rm UV}$ {\it and} with a (relatively) large \fesclya \hs (and hence \fescion), where the precise range in $M_{\rm UV}$ and \fesclya\hs both depend on the minimum Ly$\alpha$ luminosity and Ly$\alpha$ EW of the LAE sample of interest. In addition, the contribution of LAEs to cosmic reionization depends sensitively on Ly$\alpha$ EW-PDF as a function of $M_{\rm UV}$, which is not well constrained, especially at faint $M_{\rm UV}$. Here, we make a more general (and robust) point that the faint-end of the LAE LF helps constrain the $M_{\rm UV}$-dependence of \fesclya\hs and by extension, \fescion.

In the next years, the number of Ly$\alpha$ emitting galaxies at $z\sim 5.7-7$ is anticipated to grow by $\sim 1-2$ orders of magnitude with surveys performed on Subaru's Hyper Suprime-Cam (HSC). Moreover, integral-field unit spectrographs such as MUSE\footnote{\url{https://www.eso.org/sci/facilities/develop/instruments/muse.html}} will enable us to detect fainter Ly$\alpha$ emitting sources and better constrain the faint-end slope of Ly$\alpha$ emitter luminosity function, and also better characterize the (sometimes spatially resolved) spectra of Ly$\alpha$ emission lines. Recent spectroscopic observations of gravitationally lensed galaxies (e.g. Treu et al. 2015, Schmidt et al. 2016, Vanzella et al. 2016b) have uncovered several (intrinsically) UV-faint galaxies with a prominent Ly$\alpha$ emission line, and/or other spectral features such as a high [O III]$\lambda$5007/[O II]$\lambda$3727 line ratio, which favor LyC escape (Huang et al. 2016, Vanzella et al. 2016b). These observations -which support the case for an enhanced contribution of UV faint galaxies to cosmic reionization- provide a preview of what will be routinely possible with the next generation of ground-based telescopes such as the European Extremely Large Telescope (E-ELT)\footnote{\url{https://www.eso.org/sci/facilities/eelt/}}, the Thirty Meter Telescope\footnote{{\url http://www.tmt.org/}}, and the Giant Magellan Telescope (GMT)\footnote{\url{http://www.gmto.org/}}.
  
{\bf Acknowledgements} We thank the Aspen Center for Physics, where part of this collaboration was initiated, and which is supported by National Science Foundation grant PHY-1066293. MD thanks the astronomy department at UCSB for their kind hospitality. MG thanks the Physics \& Astronomy department of JHU for their kind hospitality.  AV gratefully acknowledges support from the University of San Francisco Faculty Development Fund. We thank Peng Oh and Crystal Martin for helpful discussions. Finally, we thank an anonymous referee for a constructive report which helped us improve the presentation of this work.

\label{lastpage}

\appendix
\section{Model Parameters}

Table~\ref{tab:models} provides an overview of the 14 parameters that are needed to fully characterize the multi-phase media. The second row contains the fiducial value for each parameter, which was taken from Laursen et al. (2013). The third row indicates the range of values from which we randomly drew model parameters.
\begin{table}
  \centering
  \caption{Overview of the model parameters.}
\begin{tabular}{l|rrr}
\hline
  Parameter   & Fiducial value & Allowed range & Units \\ \hline\hline
$v_{\infty,\,{\rm cl}}$ & $100.0$ & [$0.0$, $800.0$] & $\,{\rm km\,s}^{-1}$ \\
$r_{\rm {\rm cl}}$ & $100.0$ & [$30.0$, $200.0$] & $\,{\rm pc}$ \\
$P_{\rm {\rm cl}}$ & $0.35$ & [$0.0$, $1.0$] &  \\
$H_{\rm em}$ & $1000.0$ & [$500.0$, $3.0\times 10^{3}$] & $\,{\rm pc}$ \\
$f_{\rm {\rm cl}}$ & $3.5$ & [$0.8$, $8.0$] &  \\
$T_{\rm {\rm ICM}}$${}^{\dagger}$ & $10^{6}$ & [$3.0\times 10^{5}$, $5.0\times 10^{7}$] & $\,{\rm K}$ \\
$n_{\rm HI,\,{\rm ICM}}$${}^{\dagger}$ & $5.0\times 10^{-8}$ & [$10^{-12}$, $10^{-6}$] & $\,{\rm cm}^{-3}$ \\
$\sigma_{\rm i}$ & $50.0$ & [$5.0$, $100.0$] & $\,{\rm km\,s}^{-1}$ \\
$\zeta_Z$${}^{\dagger}$ & $0.01$ & [$10^{-4}$, $0.1$] &  \\
$T_{\rm {\rm cl}}$${}^{\dagger}$ & $10^{4}$ & [$5.0\times 10^{3}$, $5.0\times 10^{4}$] & $\,{\rm K}$ \\
$\beta_{\rm cl}$ & $1.5$ & [$1.1$, $2.5$] &  \\
$Z_{\rm {\rm cl}}$${}^{\dagger}$ & $0.2$ & [$0.03$, $1.0$] & $\,Z_{\odot}$ \\
$\sigma_{\rm {\rm cl}}$ & $40.0$ & [$5.0$, $100.0$] & $\,{\rm km\,s}^{-1}$ \\
$n_{\rm HI,\,{\rm cl}}$${}^{\dagger}$ & $0.35$ & [$0.03$, $3.0$] & $\,{\rm cm}^{-3}$ \\
\hline
\end{tabular}
\tablecomments{Variables marked with ${}^{\dagger}$ were drawn from a uniform distribution in log-space.}

\vspace{.5cm}
  \label{tab:models}
\end{table}

\end{document}